\documentstyle[12pt]{article}
\title{\Large \bf de Broglie oscillation, rest mass and inertia }
\author{Farhad Darabi\thanks{e-mail:  f-darabi@cc.sbu.ac.ir}\\
{\small Department of Physics, Shahid Beheshti University, Evin,
Tehran 19839,  Iran.}\\
{\small Department of Physics, Tarbiyat Moallem University, Tabriz,
Iran .}}
\begin{document}
\maketitle  
\begin{abstract}
The model of a relativistic free massive point particle is investigated in
the
context of a Hamiltonian constraint system.
It is shown that de Broglie oscillation, rest mass and inertia may be
described
within this model of Hamiltonian constraint system.
\end{abstract}  
\vspace{2cm}
\section{Introduction}

Newton's classical mechanics and gravitational theory underwent remarkable
changes after the advent of Einstein's special and general theories of relativity.
In these theories a 4-dimensional picture of our physical world was presented
providing deep insight of man about the nature of physical laws.
In this respect, a 4-dimensional description appeared as the key feature in
investigation of the fundamental physical principles. It turned out that the 4-dimensional
analysis of some physical phenomena may give the precise
3-dimensional descriptions of our experiences about them.
The basic element in this 4-dimensional analysis was the abstract object called
{\em world line element}.
The invariance property of this object under Lorentz
transformations or general coordinate transformations played the key role in
developing special and general theories of relativity.

Regarding the key role of the world line described above it is appealing to
investigate more on this object in a rather abstract way.
In this respect we are looking for some global features associated with the
world line which may emerge due to an invariance symmetry called {\em reparametrization invariance}.
We pay attention to the interesting role of
this symmetry in investigation of the two mysteries in modern physics, namely
{\em de Broglie's periodic phenomenon (oscillation)} and {\em the inertial properties
of matter}.

Louis de Broglie had presented the central point of his hypothesis on waves of
matter in a remarkable periodic phenomenon by the famous Einstein-de Broglie formula $\hbar \omega_0=m_0 c^2$,
where $\omega_0$ as the natural frequency of this phenomenon is related to the rest
mass $m_0$ of the particle. Indeed, as is well-known, this phenomenon leads,
through Lorentz transformation of the coordinate system, to de Broglie's {\em waves of matter}.
Hence the entire quantum mechanics stands on de Broglie oscillation and any
attempt to find some insight into what stands behind this phenomenon is of particular interest.

On the other hand, the issue of inertia as an {\em unresolved mystery in modern physics}
has recently been the subject of intense investigations.
Vigier \cite{VR}, used the Dirac vacuum to explain, alternative to Machian viewpoint
, the origin of inertia as a necessary consequence of the real particle motions
described by the Einstein-de Broglie-Bohm (E.d.B.B) formalism of quantum mechanics.
Recently, Rueda $et\:al.$ in a series of papers
\cite{R} have explained the inertia as scattering-like process
of the {\em zero point field} (ZPF) radiation subject to the electromagnetic vacuum.
They showed that the scattering of the incoming ZPF flux by the fundamental particles
(quarks and leptons) within the object generates a reaction force which
may account, at least in part, for inertia. Moreover, the inertial mass of an object
is that fraction of the energy of the ZPF radiation enclosed within the object that
interacts with it. More recently, they suggested that this interaction takes place at
the Compton frequency of the particle and the inertial mass of the electron is then
the reaction force due to resonance scattering of the ZPF at that frequency.
Furthermore, they suggested that de Broglie oscillation is due to a resonant
interaction with the ZPF, hence they were able to interrelate the two above mentioned
mysteries.

In this paper we take an alternative approach to study and interrelate both
of two mysteries. We seek a Hamiltonian constraint structure originating from the point particle's
action constructed by the {\em world line element}.
As is well known, the action of a relativistic free massive point particle has a
Hamiltonian constraint structure \cite{Dirac}, \cite{G} leading to the mass-shell condition,
where the space-time  coordinates are assumed to be as dynamical variables.
Indeed, this constraint structure is independent of the space-time dimensions,
so one may be allowed to reconstruct an equivalent
structure without the need to use the space-time coordinates as dynamical
variables. The motivation to follow this approach is to derive some valuable
information about a constraint dynamics which induces the local dynamics of the point
particle in space-time.

We use the {\em world line} of a relativistic free massive point particle as one
dynamical variable and parametrize its action by a world line parameter.
We then derive the associated constraint structure.
This provides a framework to justify the possible origins
of {\em de Broglie oscillation, rest mass} and {\em inertia}.

de Broglie oscillation
and the rest mass may naturally emerge due to the quantum constraint associated
with one first class constraint (in the terminology of Dirac) appearing in the model.
Indeed, using the Einstein-de Broglie formula, this quantum constraint becomes
equivalent to an eigenvalue equation whose eigenstate $e^{i \omega_0 \tau}$ corresponds to
de Broglie's localized oscillating field and
its energy $\hbar \omega_0$ corresponds to the rest mass of the point
particle ($\omega_0$ is the Compton frequency).
It is shown that what plays the role behind de Broglie oscillation and the
rest mass may be this quantum constraint which
is imposed on the rest point particle.

On the other hand, since the mass-shell condition as the 4-dimensional analogue
of our desired first class constraint holds at all times
it leads to derivation of the relativistic equation of motion and then a force in
the form $-\frac{dP_i}{d\tau}$ is interpreted physically as {\em inertia}.
Although the physical mechanism behind this interpretation may be the ZPF scattering
\cite{R}, but the {\em causal} origin of this
reaction force seems to be in the reparametrization invariance of the point particle system which
may be induced as a property resulting from the same invariance property in large universe
scale. In this approach, the local inertial properties of the point particle are generated
due to the ``{\em timeless}'' property of the universe and its causal
structure. {\em Inertia of the particle then becomes a direct evidence indicating that
no absolute time exists in the universe.}

\section{Constraint System}

In this section we will use Dirac's formalism of Hamiltonian constraint systems \cite{Dirac}.
We start with the action of a relativistic free massive point particle
\begin{equation}
I = - m_0 c \int_{s_1}^{s_2}\!d{\bf s}
\end{equation}
where $m_0$ is the rest mass of the particle, $c$ is the velocity of light
and $d{\bf s}$ is the world line element.
We now parametrize the action as
\begin{equation}
I = - m_0 c \int_{\tau_1}^{\tau_2}\!\dot{{\bf s}} d\tau
\end{equation}
where $\dot{{\bf s}} = \frac{d{\bf s}}{d\tau}$ and $\tau$ is the world line parameter
usually taken as proper time.
As mentioned in the introduction, we have taken ${\bf s}$ as the only dynamical variable
in order to find some global
characteristics of the model.
The linear action (2) admits a Hamiltonian constraint structure with the primary constraint
\begin{equation}
\phi \equiv P_s + m_0 c\approx 0
\end{equation}
where $P_s$ is the momentum conjugate to ${\bf s}$.
The original Hamiltonian $H_0$ is zero for the linear action, so the
total Hamiltonian is given
\begin{equation}
H_T = \lambda(\tau) \: \phi
\end{equation}
where $\lambda(\tau)$ is an arbitrary function of time. Hence the dynamics is fully
controlled by the constraint (3). The consistency condition
\begin{equation}
0 \approx \dot{\phi} = \{\phi,H_T\}
\end{equation}
is automatically satisfied since $\{\phi,\phi\} \approx 0$, so that there is
no secondary constraints.
Now, the constraint (3) is by definition \cite{Dirac} a first class
constraint which generates the following infinitesimal local gauge transformations
\begin{equation}
\delta {\bf s} = \epsilon(\tau) \{{\bf s},\phi\} = \epsilon(\tau)
\end{equation}
\begin{equation}
\delta P_s = \epsilon(\tau)\{P_s,\phi\} = 0
\end{equation}
where $\epsilon(\tau)$ is an infinitesimal arbitrary function of time.
It follows from (6) that the variable ${\bf s}$ has a gauge orbit. On the other
hand, from (7) we find that the momentum $P_s$ is a gauge invariant
quantity on this orbit and should have a physical content.
The equations of motion are
\begin{equation}
\dot{{\bf s}} = \{{\bf s},H_T\} = \lambda(\tau)
\end{equation}
\begin{equation}
\dot{P_s} = \{P_s,H_T\} = 0
\end{equation}
which integrate to
\begin{equation}
{\bf s}(\tau) = \int^{\tau}\!\lambda(\tau')\:d{\tau'}
\end{equation}
\begin{equation}
P_s = \mbox{Const.}
\end{equation}
These are consistent with (6) and (7) provided that $\delta \lambda =
\dot{\epsilon}(\tau)$.
With the identification
$$
\epsilon(\tau)=\lambda(\tau) \eta(\tau)
$$
where $\eta(\tau)$ is defined for an infinitesimal world line parametrization as
$$
\eta(\tau)=\tau -\tilde{\tau}(\tau)
$$
one can see that, using equations of motion (8) and (9), the infinitesimal
local gauge transformations (6), (7) correspond to the world line reparametrizations.
The freedom in the choice of Lagrange multiplier $\lambda(\tau)$ corresponds to
the freedom in the choice of world line parametrization. Hence, choosing the parametrization
means the gauge-fixing of the system through a choice for $\lambda(\tau)$. However,
different functions $\lambda(\tau)$ leading to the same value of the
action $I$ lie on the same
gauge orbit and describe identically the same physical situation.
\footnote{
Note that as a result of equation (8) and $\delta \lambda=\dot{\epsilon}(\tau)$ with $\epsilon(\tau_i)=0$
we have
$$
\delta I=-m_0 c \:\delta \int_{\tau_1}^{\tau_2} \lambda(\tau) d\tau=0.
$$
Therefore, the action $I$ is the {\em Teichm\"{u}ller} parameter labelling the gauge
orbit of the system. For positive energy with $\lambda>0$ we have forward propagation
in time for $I<0$ (a particle) and backward propagation for $I>0$ (an antiparticle) \cite{G}. }

\section{de Broglie oscillation and rest mass}

The constraint (3) as the heart of this model merits considerable attention. We know
from special theory of relativity the well-known relation
\begin{equation}
E^2=P^2 c^2+m_0^2 c^4
\label{*}
\end{equation}
which relates the energy content $E$ of a point particle to its spatial momentum $P$
and rest mass $m_0$. Using the constraint (3) $|P_s|=m_0c$
, we may rewrite (\ref{*}) as
\footnote{The notation $\approx$ in Eq. (3) does not mean an approximate relation.
It merely tells us that the constraint $|P_s|=m_0c$ should be imposed after derivation
of the equations of motion \cite{Dirac}.}
\begin{equation}
E^2=(P^2 +P_s^2) c^2
\label{+}
\end{equation}
reflecting the fact that energy content of a point particle emerges due to
two types of linear momentums, its usual spatial one $P$ and the intrinsic one
$P_s$ conjugate to the world line coordinate. For a particle at rest in space-time $P=0$, equation (\ref{+}) becomes
\begin{equation}
E=|P_s| c
\end{equation}
which relates the energy content of the point particle to its intrinsic
momentum $P_s$ on the world line. This is like the corresponding formula $E=Pc$ for massless particles (photons) for which we have
\begin{equation}
\hbar \omega=Pc
\end{equation}
where the angular frequency $\omega$ is related to the spatial momentum $P$ of these particles
according to {\em wave-particle duality}. Insisting on the generality of {\em wave-particle
duality} we may relate, as well,
an intrinsic angular frequency $\omega_0$ (as the {\em wave} characteristic) to the intrinsic
momentum $P_s$ (as the {\em particle} characteristic) of the rest point particle as
\begin{equation}
\hbar \omega_0=|P_s| c.
\label{1}
\end{equation}
Therefore, using $|P_s|=m_0c$, Einstein-de Broglie relation $\hbar \omega_0=m_0 c^2$
emerges naturally due to generalization of {\em wave-particle duality} to the rest particles
\footnote{de Broglie introduced the formula $\hbar \omega_0=m_0 c^2$ directly without
mentioning to the wave-particle duality as its justification. Indeed, Einstein-de
Broglie relation in the form $\hbar \omega_0=m_0 c^2$ for
a rest mass does not indicate {\em wave-particle duality} in the sense of de Broglie's
waves of matter. This is because Einstein-de Broglie relation reads as the Compton
wavelength $\lambda=\frac{\hbar}{m_0 c}$ and if one wants to describe it in the
sense of de Broglie's waves of matter then a rest particle should have the spatial
momentum $P=m_0 c$ which is a contradiction. In other words, although the wavelength
$\lambda$, in the left hand side, manifests the {\em wave} aspect but the rest mass $m_0$,
in the right hand side, by no means indicates
the {\em particle} aspect. The {\em particle} aspect of the rest particles emerges
whenever the intrinsic momentum $P_s$ is attributed to them. This gives rise to the
wave-particle duality expressed as the formula $\lambda=\frac{\hbar}{P_s}$. This formula
relates the Compton wavelength $\lambda$ to the particle's intrinsic momentum $P_s$ on the
world line coordinate. }.
Einstein-de Broglie relation as realization of mass-energy equivalence relates
the concept of rest mass to a type of
energy which appears to be the energy of massless particles namely, $\hbar \omega_0$.
Therefore, intuitively, one may think that a
massive rest particle as viewed by a rest inertial observer is nothing but a photon-like particle, moving with light velocity
$\lambda=c$ on its world line, whose energy content appears, through mass-energy
equivalence, as the particle's rest mass\footnote{Here, only the mass property of the particles are compared without referring
to their other properties such as spin or charge.}. This 
may shed light on the origin of de Broglie's oscillation $e^{-i\omega_0 \tau}$ as it
seems to be the time dependent part of a wave propagation along the world line associated with
the photon-like particle. This is because a rest observer also moves with the light velocity
$\lambda=c$ on its own world line so only the time evolving part of this wave
propagation manifests.

The {\em wave-particle duality} for a 
rest particle becomes more apparent if we try to derive de Broglie
oscillation from this constraint system. We remind that the entire quantum mechanics
stands on {\em wave-particle duality} with wave functions and eigenvalues as indicating
the wave and particle aspects respectively. So, if we can derive de Broglie oscillation itself from
quantum mechanics then we may conclude that this periodic phenomenon follows,
in principle, the {\em wave-particle duality}.
To this end, we pay attention for quantization of the present model.

Quantization of this constraint system is done by operating on the Hilbert subspace $\mid\psi>$
by the operator form of the constraint (3) as \cite{Dirac}
\begin{equation}
( \hat{P}_s + m_0c )\:\mid\psi> = 0 .
\label{Q}
\end{equation}
Using Einstein-de Broglie relation and $i \hbar \frac{\partial}{\partial \tau}=-c\frac{\hbar}{i} \frac{\partial}{\partial {\bf s}}$
for the rest particle obeying $d{\bf s}=\lambda d\tau$ with $\lambda=c$, the quantum constraint
(\ref{Q}) becomes the eigenvalue equation
\begin{equation}
i \hbar \frac{\partial}{\partial \tau} \mid\psi> = E \mid\psi>
\end{equation}
with energy and wavefunction as
$$
E=\hbar \omega_0
$$
\begin{equation}
\mid\psi(\tau)> \sim e^{-i\omega_0 \tau}.
\label{y}
\end{equation}

{\em Remarkably, the wave function (19) as the physical solution satisfying the
quantum constraint (17) is exactly what we have known as de Broglie's periodic
phenomenon and the eigenvalue $\hbar \omega_0$ is the corresponding energy}.
This wave function is the physical state describing the photon-like particle,
as is observed by the rest observer.
If we believe that quantum mechanics, as stands on de Broglie's oscillation, provides us the proper observables with physical
reality then by translating the quantum constraint (17) into an eigenvalue equation (18) we are able
to derive the rest mass $m_0$, as a physical {\em observable} $\frac{\hbar \omega_0}{c^2}$,
from de Broglie's periodic phenomenon.
In other words, the rest mass $m_0$ itself does not qualify as an observable; it
is $\frac{\hbar \omega_0}{c^2}$ which physically we observe as the rest mass
\cite{H}. 

Moreover, the quantum constraint (17) or equivalently Eq (18) seems to describe the
quantum motion well-known as {\em Schr\"{o}dinger Zitterbewegung}\cite{Ho}.

\section{Inertia}

The issue of inertia is basically correlated with the phenomenon of acceleration.
Suppose we have a rest point particle in an inertial reference frame. Actually, since the particle does not
undergo any external force the action for a free point particle (1) is then applied
and leads to the first class constraint
$$
\phi \equiv P_s + m_0 c\approx 0 .
$$
In order to find what effects such a constraint structure generates in space-time
we resort to the 4-dimensional analogue of the constraint (3)
as the mass-shell condition \cite{G}
\begin{equation}
\psi\equiv P^{\mu} P_{\mu}+m_0^2 c^4\approx 0
\label{z}
\end{equation}
subject to the total Hamiltonian
$$
H_T=\lambda(\tau) \psi
$$
where $P_{\mu}$ is the 4-momentum of the particle defined on the Minkowski space-time
with metric (- + + +).
Time independence of the mass-shell condition (\ref{z}) $\dot{\psi}=\{\psi, H_T\}=0$,
means trivially
\begin{equation}
\frac{d P_{\mu}}{d\tau}=0
\end{equation}
which is the equation of motion for a free particle where $\tau$ is the proper time.
However, we find that the following non-trivial condition
\begin{equation}
P^{\mu}\frac{dP_{\mu}}{d\tau}=0
\label{y}
\end{equation}
with $\frac{dP_{\mu}}{d\tau} \neq0$ may also be satisfied in consistency with the
first class constraint (20).
To this end, we define the generalized spatial force $Q_i$ as the
{\em force of constraint}
\begin{equation}
Q_i\equiv F_i-\frac{dP_i}{d\tau} \:\:\:\:\:i=1, 2, 3
\end{equation}
where $F_i$ indicates the spatial components of the four-force $F_{\mu}$ and
$-\frac{dP_i}{d\tau}$ is assumed to be a force with physical origin in the same foot
as $F_i$.
The physical nature of the force $-\frac{dP_i}{d\tau}$ may be justified according
to the recent results obtained by Rueda {\em et.al.}
in that a reaction (physical) force $\stackrel{\rightarrow}{F}_r=\frac{d\stackrel{\rightarrow}{P}_r
}{d\tau}=-\frac{d\stackrel{\rightarrow}{P}}{d\tau}$ (generated by ZPF scattering)
is interpreted as inertia so that one can rewrite equation (23) as $Q\equiv \stackrel{\rightarrow}{F}
+\stackrel{\rightarrow}{F}_r$ \cite{R}.
It is easy to show that $Q_i=0$ satisfies the condition (\ref{y}). To see, we rewrite
(\ref{y}) as
\begin{equation}
E\frac{dE}{d\tau} -P_i\frac{dP_i}{d\tau}c^2=0
\end{equation}
which leads to
\begin{equation}
F_i=\frac{dP_i}{d\tau}
\label{x}
\end{equation}
where
$$
\frac{dE}{d\tau}=F_i V_i\:\:\:,\:\:\:E=\gamma m_0 c^2\:\:\:and\:\:\:P_i=\gamma m_0 V_i
$$
have been used and $\gamma$ is the relativistic factor \cite{Gold}
\footnote{Here, the four-momentum denoted by $K_{\mu}$
in \cite{Gold} is exchanged symbolically by $F_{\mu}$ for convenience.}.
Therefore, the vanishing of the force of constraint $Q_i$ implies the mass-shell condition to be hold
at all times.
The interpretation of vanishing $Q_i$ is that the acceleration, as well as uniform motion,
is also consistent with this first class constraint such that
the active force $F_i$ is balanced, according to Newoton's third
law, by an equal and opposite directional (physical)
force $-\frac{dP_i}{d\tau}$ which we shall interpret it as
inertia.
As far as we concern with the present constraint system it is reasonable
to say that during the acceleration of particle a reaction force
against the agent of acceleration is generated as the price to be paid to preserve the
constraint (20) which defines, on the other hand, the causal structure of space-time in the model.

As is usual in the study of non-inertial frames in classical mechanics,
we may also transfer the term $-\frac{dP_i}{d\tau}$ to the right hand side of equation $Q_i=0$
to obtain the equation (\ref{x}) which is interpreted as the relativistic
equation of motion. These interpretations will be reasonable if we apply
the general law of motion in the form
\begin{equation}
\sum \mbox{Physical Force} = \frac{d}{d\tau} (\mbox{Particle Momentum})
\end{equation}
in both inertial and accelerated (co-moving) frames of reference.
Then the equation
\begin{equation}
F_i-\frac{dP_i}{d\tau}=0
\label{w}
\end{equation}
compared to (26) implies that in the left hand side the agent of particle dynamics, namely the
physical motive force $F_i$ co-moving with the particle experiences an equal and opposite directional
{\em physical} force $-\frac{dP_i}{d\tau}$ as the inertia which results in
the right hand side a zero acceleration of particle in the co-moving frame.
A realistic example for this situation is to suppose an accelerated rocket to which a particle
has been attached by an string. In the co-moving frame, the rocket will then experience the inertia of
particle as a reaction force against the active force imposed by string, but no acceleration is attributed to the particle simply
because it moves in contact with the rocket.
On the other hand, the equation written as
$$
F_i=\frac{dP_i}{d\tau}
$$
compared to (26) implies that the physical force $F_i$ (in the left hand side) produces a dynamics for the particle
according to $\frac{dP_i}{d\tau}$ (in the right hand side) in the inertial frame and expresses the relativistic
equation of motion.

Equations (25) and (27) may be also interpreted as expressing the circular motion
of the point particle in both inertial and rotating frames of reference, respectively.
In this respect, $F_i$ would be the centripetal force and the term $-\frac{dP_i}{d\tau}$
would indicate the centrifugal force in the rotating frame.

We conclude that in the context of present paper {\em inertia as the reaction force against acceleration may
emerge due to resistance of the first class constraint to breakdown}. The breakdown
occurs either when an absolute time parameter exists or when the causal structure of space-time 
breaks down. Therefore, it may be said that the absence of an absolute time parameter or, 
equivalently, the causal structure described by the mass-shell condition in this system is the possible
conceptual origin of the inertia. On the other hand, we know from cosmology that
the universe, as a whole, may be described by a similar constraint
structure as here with a first class constraint known as Wheeler-DeWitt equation.
This equation also implies that there is no absolute time parameter external to the
universe.

For an interested Machian, it would be appealing to relate the whole universe and the inertia
of particle through the general property of the universe:``{\em There is
no absolute time}''.
In this regard, the inertia of a particle emerging from reparametrization
invariance of its action may be induced locally by this global
cosmological property. This is because it is not reasonable to have no absolute time in cosmological scale
but have an absolute time in local scales.

The old question that ``{\em How a material object understands its motion to manifest
the inertial effects?} '' was answered by Newton as: ``{\em It feels the action of absolute
space}'', and by Mach as: ``{\em It feels the action of cosmic matter}''.
Another answer, based on this
approach, may be given as:
``{\em It knows the absence of absolute time and feels the action of space-time causal structure}''. Therefore, the universal property of
reparametrization invariance seems to be a missed chain which
links the universe to the local particle dynamics without the need to {\em action at distance} violating
the causality principle. In simple words, the state of being at rest, in uniform motion
and in accelerated motion of a particle are correlated with the state of universe
through this universal property. In this regard, the first and second laws of Newton
seem to be two different manifestations of this universal property ( see equations (21) and (25) ).
Although this approach to the issue of inertia is not fully Machian
\footnote{Wheeler-DeWitt equation is also valid for
an empty universe. Therefore, in the context of present paper the particle may have inertia even in an empty universe.}
(as there is no direct role playing by the
matter content of the universe)
but, unlike the Machian viewpoint, has the advantage of being
compatible with general relativity as a realization of the reparametrization invariant
theory. If correct, this approach would substitute for Mach's principle and imply
that the inertial effects are pure relativistic in nature preserving the causality
in general agreement with \cite{R}.

\end{document}